\documentclass[journal,twoside,web]{ieeecolor}
\usepackage{tmi}
\usepackage{cite}
\usepackage{amsmath,amssymb,amsfonts}
\usepackage{algorithmic}
\usepackage{graphicx}
\usepackage{textcomp}
\usepackage{multirow}
\usepackage[table,xcdraw]{xcolor}
\def\BibTeX{{\rm B\kern-.05em{\sc i\kern-.025em b}\kern-.08em
    T\kern-.1667em\lower.7ex\hbox{E}\kern-.125emX}}
\begin{document}
\title{Towards Collective Intelligence: Uncertainty-aware SAM Adaptation for Ambiguous Medical Image Segmentation}


\author{Mingzhou Jiang,
        Jiaying Zhou,
        Junde Wu,~\IEEEmembership{Graduate Student Member, IEEE},
        Tianyang Wang,
        Yueming Jin,~\IEEEmembership{Member, IEEE},
        and~Min Xu,~\IEEEmembership{Member, IEEE}
\thanks{Manuscript received 17 November 2025. \textit{(Mingzhou Jiang, Jiaying Zhou and Junde Wu contributed equally to this work.) (Corresponding author: Min Xu)}}
\thanks{Mingzhou Jiang and Jiaying Zhou are with the Department of Computer Science, The University of Alabama at Birmingham, Birmingham, AL 35294, United States (e-mail: 011122zmj@gmail.com; jiayingzhou2127@outlook.com).}
\thanks{Junde Wu is with the Doctoral Training Centre, University of Oxford, Oxford OX1 2JD, U.K., also with Department of Biomedical Engineering and Department of Electrical and Computer Engineering, National University of Singapore, Singapore 119276 (e-mail: jundewu@ieee.org).}
\thanks{Tianyang Wang is with the Department of Computer Science, The University of Alabama at Birmingham, Birmingham, AL 35294, United States (e-mail: tw2@uab.edu).}
\thanks{Yueming Jin is with Department of Biomedical Engineering and Department of Electrical and Computer Engineering, National University of Singapore, Singapore 119276 (e-mail: ymjin@nus.edu.sg).}
\thanks{Min Xu is with the Computational Biology Department, Carnegie Mellon University, Pittsburgh, PA 15213, United States, and also with Mohamed bin Zayed University of Artificial Intelligence, Abu Dhabi 999041, United Arab Emirates (e-mail: mxu1@cs.cmu.edu).}}
\maketitle

\begin{abstract}
Collective intelligence from multiple medical experts consistently surpasses individual expertise in clinical diagnosis, particularly for ambiguous medical image segmentation tasks involving unclear tissue boundaries or pathological variations. The Segment Anything Model (SAM), a powerful vision foundation model originally designed for natural image segmentation, has shown remarkable potential when adapted to medical image segmentation tasks. However, existing SAM adaptation methods follow a single-expert paradigm, developing models based on individual expert annotations to predict deterministic masks. These methods systematically ignore the inherent uncertainty and variability in expert annotations, which fundamentally contradicts clinical practice, where multiple specialists provide different yet equally valid interpretations that collectively enhance diagnostic confidence.
We propose an Uncertainty-aware Adapter, the first SAM adaptation framework designed to transition from single expert mindset to collective intelligence representation. Our approach integrates stochastic uncertainty sampling from a Conditional Variational Autoencoder into the adapters, enabling diverse prediction generation that captures expert knowledge distributions rather than individual expert annotations. We employ a novel position-conditioned control mechanism to integrate multi-expert knowledge, ensuring that the output distribution closely aligns with the multi-annotation distribution. Comprehensive evaluations across seven medical segmentation benchmarks have demonstrated that our collective intelligence-based adaptation achieves superior performance while maintaining computational efficiency, establishing a new adaptation framework for reliable clinical implementation.

\end{abstract}


\begin{figure}
    \centering
    \includegraphics[width=0.95\linewidth]{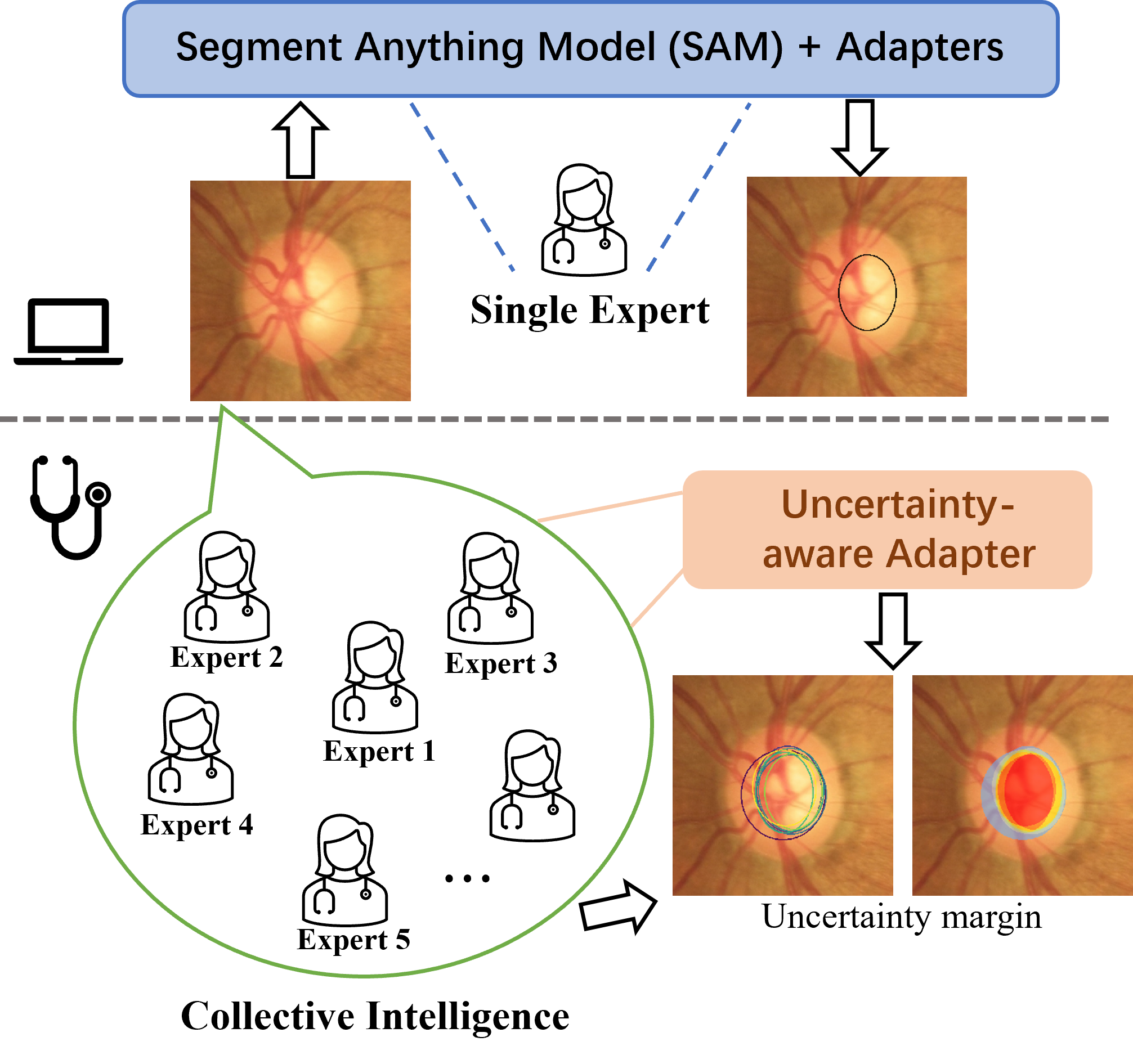}
    \caption{An example of previous SAM adaptation methods (based on single-expert strategy) and our proposed \textbf{Uncertainty-aware Adapter} based on multi-expert collective intelligence for clinical scenarios}
    \label{fig1}
\end{figure}
\section{Introduction}
\label{sec:introduction}
\IEEEPARstart {V}{ision} foundation models such as the Segment Anything Model (SAM) \cite{kirillov2023segment}, pretrained on large-scale natural image datasets, have revolutionized computer vision with remarkable generalization capabilities. Specifically, SAM employs the interactive segmentation framework and delivers outstanding performance across multiple downstream domain segmentation tasks \cite{zhang2024uv,zhang2024fantastic}, and has swiftly become the darling of natural image segmentation. Many researchers have proposed to adapt SAM to medical image segmentation, positioning it as a specialist for multiple medical domains \cite{ma2024segment, deng2023segment, gao2023desam}. A popular effective approach is the adapter technique, which involves inserting a bottleneck module with only a few parameters into the model, thus retaining its superior performance while maintaining efficiency \cite{wu2023medical, cheng2023sam, chen2023sam}.

However, while SAM achieves impressive performance when adapted to medical segmentation through efficient adapter techniques, these adaptations remain confined to a single expert paradigm. This paradigm learns from a single annotation and generates deterministic outputs, which misaligns with medical practice. In routine clinical scenarios, the fundamental challenge emerges from two sources of inherent uncertainty: ambiguous tissue and organ boundaries in medical images, and reasonable differences in physician experience and clinical judgment \cite{fu2020retrospective, joskowicz2019inter}. Annotations from individual physicians can contain errors arising from complex cases or human oversight, yet current AI systems commonly rely on these single expert annotations without addressing such inherent limitations \cite{fu2020retrospective}. By contrast, clinical practice typically requires collecting and synthesizing annotations from multiple experts to maintain safety standards, leveraging collective intelligence to enhance diagnostic reliability \cite{karimi2020deep, liao2022learning, wu2024diversified}. Fig.\ref{fig1} shows an example of clinical diagnosis of glaucoma. In the collective intelligence paradigm, when segmenting the optic cup, senior physicians may provide more aggressive boundaries while younger specialists might offer more conservative delineations \cite{erick2024uncertainty}. Each expert perspective contains valuable clinical insights that would be lost if current SAM adaptation methods rely on a single expert. 

\begin{figure}
    \centering
    \includegraphics[width=\linewidth]{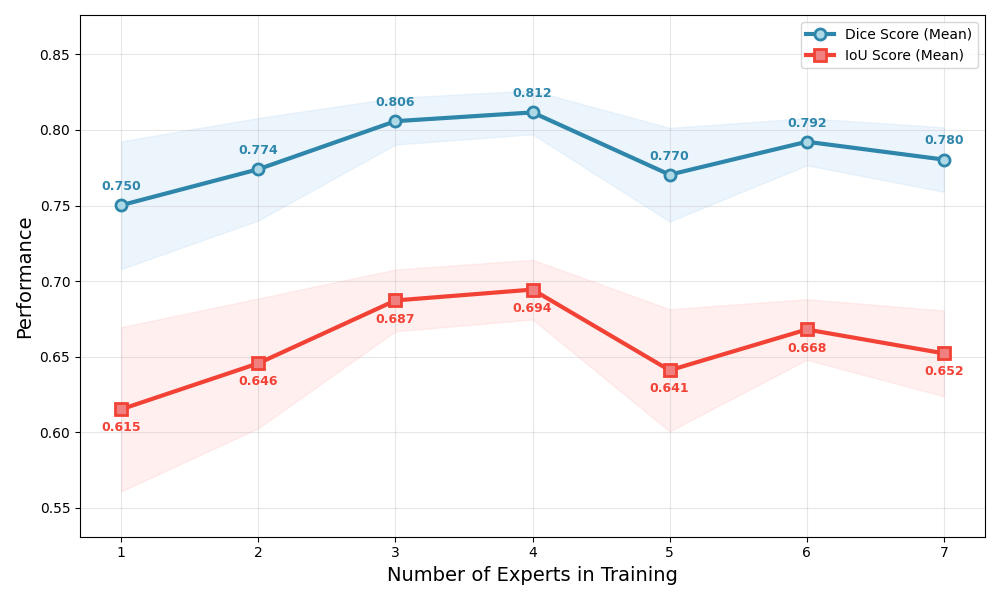}
    \caption{The impact of aggregating different numbers of expert annotations for adapting SAM to optic cup segmentation.}
    \label{fig2}
\end{figure}

Incorporating multi-expert perspectives using methods such as majority voting can calibrate individual expert annotation errors and is typically used as the gold standard for training and evaluation \cite{schaekermann2019understanding, ji2021learning}.
Nevertheless, as illustrated in Fig.\ref{fig2}, we discovered that multi-expert consensus can lead to complex annotation patterns, thereby complicating the learning process for SAM adaptation methods. For the analysis, we systematically increased the number of experts used to generate consensus masks for training SAM adapters on optic cup segmentation, then evaluated all models on the final consensus mask generated by all experts. It is evident that as the number of participating experts increases, performance first rises and then declines, followed by a plateau phase. This reveals the inherent limitations of the previous SAM adaptation methods, which can only learn specific annotator information and cannot effectively capture multi-expert knowledge.

Therefore, effectively adapting SAM to leverage the advantages of multi-expert collective intelligence while maintaining computational efficiency and interactive capabilities is worth exploring. By transforming SAM from one-to-one deterministic mapping to one-to-many diverse plausible mapping, we can transition from a single-expert paradigm to collective intelligence, thus enhancing the model's clinical applicability. On the other hand, considering that SAM serves as an interactive segmentation framework, the impact of visual prompts cannot be ignored \cite{chen2023sam, ma2024segment}. Particularly, when segmenting medical images with unclear boundaries, different prompts significantly affect model performance. However, most adaptation methods only perform limited interaction of prompt embeddings in the decoder, and the lack of user prompt awareness would diminish SAM's prompt sensitivity, leading to subpar performance after fine-tuning \cite{wu2023medical}.

Driven by these problems, we propose an Uncertainty-aware Adapter, a user-prompt-sensitive module that efficiently transforms vanilla SAM from a single-expert paradigm to collective intelligence representation. The high-level idea is to incorporate the probabilistic sampling process into adapters, allowing the model to generate diverse predictions that simulate the collective intelligence of multiple experts. Inspired by Bayesian Neural Networks \cite{kohl2018probabilistic, Bhat_2023, viviers2023probabilistic}, we customize a latent space for sampling to represent the range of valid expert perspectives and propose a position-conditioned strategy to integrate this expert diversity into the adapters. Specifically, we propose a lightweight attention mechanism called Position-conditioned Attention (Poscon-Att). Poscon-Att establishes latent relationships between stochastic uncertainty sampling and adapter insertion positions within SAM's hierarchical architecture, thereby unlocking SAM's capability to represent segmentation uncertainty. Furthermore, we introduce Prompt Channel Attention (Prompt-CAtt) to incorporate visual prompts into the Uncertainty-aware Adapter. In Prompt-CAtt, we use prompt embeddings to generate a series of channel attention weights that can be efficiently applied to the adaptation embedding, thereby facilitating extensive and in-depth visual prompt interactions. This paradigm shift enables SAM to generate diverse predictions close to multi-expert distributions rather than singular outputs, achieving collective intelligence representation while preserving computational efficiency through parameter-efficient adaptation.
Our contributions can be summarized as follows:

• We identify the limitations in current SAM adapter approaches and propose a novel collective intelligence adapter paradigm, which can capture multi-expert knowledge distributions and enhance clinical applicability.

• We propose an Uncertainty-aware Adapter framework that employs parameter-efficient adapter modules to adapt SAM with collective intelligence capability through two novel designs: Position-conditioned Attention employs adapter position information as a condition to control multi-expert knowledge learning; Prompt Channel Attention integrates visual prompts into adapter fine-tuning while preserving multi-expert knowledge synthesis and maintaining prompt sensitivity.

• We evaluate our model, Uncertainty-aware SAM (UA-SAM), on seven ambiguous medical segmentation benchmarks. Our model demonstrates strong uncertainty representation capability with superior user-prompt adaptability after parameter-efficient fine-tuning, achieving state-of-the-art performance on all benchmarks.

\begin{figure*}[!t]
    \centering
    \includegraphics[width=\linewidth]{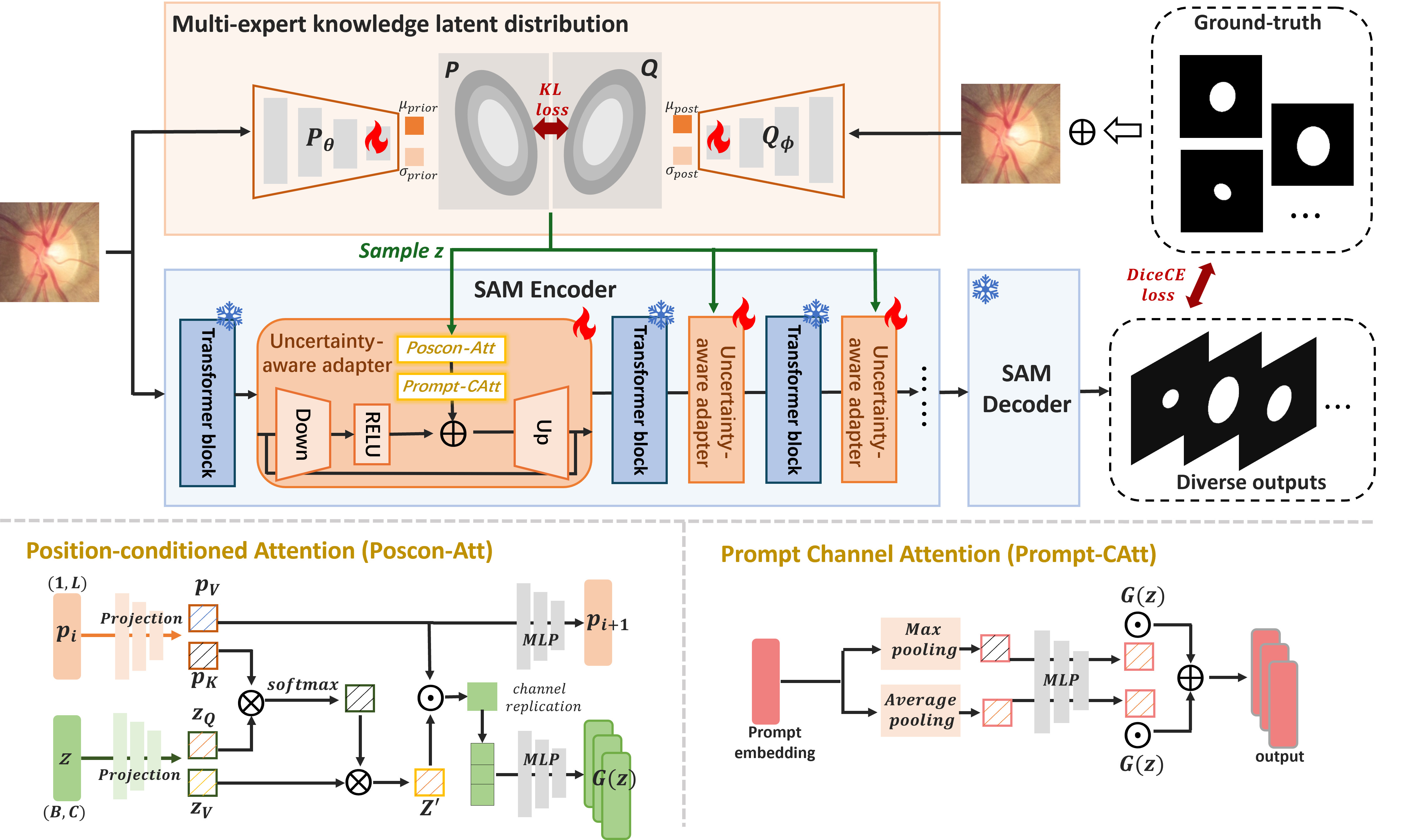}
    \caption{Overview architecture of framework UA-SAM. We froze the parameters of SAM and only updated the Uncertainty-aware Adapter parameters. Note that we did not show the prompt encoder.}
    \label{fig3}
\end{figure*}

\section{Related Works}
\subsection{SAM Adaptation: the single expert paradigm.} The Segment Anything Model (SAM) \cite{kirillov2023segment},
a groundbreaking framework for interactive segmentation,
has demonstrated significant impact in natural image segmentation. 

However, due to lack of domain-specific medical knowledge, SAM shows limited performance in medical image segmentation tasks. Various adaptation strategies have been explored \cite{chen2023sam, ma2024segment, liu2023adapting, yan2024after, chen2024ma}. Among these, adapter-based techniques, inspired by advances in natural language processing \cite{pmlr-v97-houlsby19a}, have proven particularly efficient by enabling fine-tuning with minimal additional parameters while maintaining strong performance.
Wu et al. \cite{wu2023medical} introduced the Medical SAM Adapter (Med-SA), a lightweight yet effective approach that enhanced SAM's medical segmentation capabilities by updating only 2 percent of the total parameters. They further proposed HyP-Adpt to enable prompt-conditioned adaptation, recognizing the critical role of user-provided prompts in medical image segmentation. Similarly, Cheng et al. \cite{cheng2023sam} proposed SAM-Med2D, which incorporated learnable adapter layers in each Transformer block to acquire domain-specific medical knowledge, achieving accurate segmentation of multiple organs, tumors, and lesions. Adapter-based tuning has recently been carried over to the newer model SAM2 \cite{ravi2024sam}, yet its performance on medical image segmentation remains limited \cite{sengupta2024sam}. Despite these advances, a fundamental limitation is that these approaches operate within a single-expert paradigm, treating medical image segmentation as a deterministic task. This deterministic output tends to produce over-confident predictions that fail to capture the inherent ambiguity in medical images.

\subsection{Medical Uncertainty Modeling: multi-expert collective intelligence.}
Due to ambiguous boundaries of lesions, tissues, and organs, along with variations in clinical expertise, the same medical image can lead to different annotations \cite{sylolypavan2023impact}. This inherent ambiguity necessitates uncertainty-aware methods that can generate multiple plausible segmentation hypotheses \cite{ji2021learning}, which hold greater clinical value than deterministic approaches.
To address this challenge, uncertainty quantification methods such as model ensembles \cite{guan2018said} and variational inference \cite{kohl2018probabilistic} have been proposed to capture output diversity. A notable work is Prob U-Net \cite{kohl2018probabilistic}, which combined U-Net with a conditional variational autoencoder (CVAE) to estimate aleatoric uncertainty from multi-annotator data. Unlike traditional approaches that produce pixel-wise probabilities, Prob U-Net generated consistent segmentation maps with calibrated likelihoods, effectively capturing output uncertainty and rare modes. This CVAE-based uncertainty modeling has inspired subsequent works. Bhat et al. \cite{Bhat_2023} analyzed how different latent space distributions affect the model's ability to capture annotation variability in lung tumors and white matter hyperintensities. Viviers et al. \cite{viviers2023probabilistic} extended to 3D volumetric data and incorporated Normalizing Flows to enable more flexible modeling of complex output distributions.

However, these uncertainty quantification methods are primarily built upon U-Net architectures, which have inherent limitations in capturing long-range dependencies and generalizing across diverse medical imaging tasks. Moreover, the paradigm shift of SAM adaptation methods from single expert approaches to multi-expert collective intelligence modeling remains underexplored, representing a critical gap between current adaptation strategies and clinical practice requirements.

\section{Methods}
Conventional SAM adaptation methods employ deterministic processing where, given an input image $X$ and prompts $p$, the adapted model produces a deterministic segmentation $M = \text{SAM}_{\text{adapter}}(X, p)$. This one-to-one mapping limits clinical applicability and misaligns with real-world clinical scenarios. In practice, clinical applications rely on a multi-expert collective intelligence paradigm where preserving and synthesizing diverse expert perspectives is essential for reliable diagnosis. To bridge this gap, we propose UA-SAM, which upgrades SAM from a single-expert paradigm to a collective-intelligence paradigm through a lightweight Uncertainty-aware Adapter.

\subsection{Overall architecture}
The core challenge in achieving collective intelligence representation lies in enabling SAM to generate diverse, close to the actual distribution of multi-expert knowledge while maintaining computational efficiency and interactive capabilities. 
To address it, we propose UA-SAM, a 
collective intelligence adaptation paradigm, the overall framework is shown in Fig.~\ref{fig3}. Firstly, we explicitly formulate the probabilistic distribution $\mathcal{M}e(z|X)$ of \textbf{M}ulti-\textbf{e}xpert knowledge using conditional Bayesian modeling \cite{kohl2018probabilistic, Bhat_2023, viviers2023probabilistic}, where any sample $z$ drawn from the distribution is regarded as an independent insight of different experts, corresponding to a plausible segmentation result. 
Subsequently, a novel Uncertainty-aware Adapter that is plugged into SAM, enabling it to learn multi-expert annotation pattern from the distribution $\mathcal{M}e(z|X)$. The adapter employs conditional interaction mechanism $cond(\cdot)$ to learn the mapping from low-dimensional vector $z$ to segmentation. This capacitates the fine-tuned SAM to produce diverse segmentation outputs $\tilde{M}$ for the same input $X$ by sampling different $z$, as formulated in Eq.\ref{eq1}
\begin{equation}
    \tilde{M} = \text{UA-SAM}(X, cond(z)), \quad st. \ z \sim \mathcal{M}e(z|X)\label{eq1}
\end{equation}

\subsection{Multi-expert knowledge latent distribution}
To generate the distribution $\mathcal{M}e(\cdot)$ representing multi-expert annotations knowledge, it is imperative to estimate the parameters $\Theta$ of this distribution. We adopt a diagonal Gaussian distribution $\mathcal{N}(\mu, \text{diag}(\sigma))$ ($\mu,\sigma \in {\mathbb{R}}^{B\times C}$, where $B$ is the batch size and $C$ is the latent distribution dimension) to characterize the set of stochastic samples $z$. The distribution is learned by minimizing the KL divergence loss function (in Eq.\ref{e4}) between a Prior network $P_{\theta}$ and a Posterior network $Q_{\phi}$.
The Prior network $P_{\theta}$ takes only the raw image $X$ and learns the mapping from the image to the multi-expert annotations space. The Posterior network $Q_{\phi}$ employs a label-sampling strategy \cite{jensen2019improving}, where it randomly selects one annotation $Y$ from the candidate multi-expert labels, concatenating it with the raw image, and using this combined input to supervise the prior representation. Prior network $P_{\theta}$ and Posterior network $Q_{\phi}$ are implemented using a 4-layer convolutional network. During training, we represent the knowledge latent distribution $\mathcal{M}e(\cdot)$ with the posterior distribution $Q$, whereas during inference, we rely on the prior distribution $P$. We formulate the workflow by the following equations:
\begin{equation}
    \mathbf{z_{infer}} \sim \mathcal{P}(\cdot | \mathbf{X}) = \mathcal{N}(\mu_{\text{prior}}(\mathbf{X}), \text{diag}(\sigma_{\text{prior}}(\mathbf{X})))\label{e2}
\end{equation}
\begin{equation}
    \mathbf{z_{train}} \sim \mathcal{Q}(\cdot | \mathbf{X}, \mathbf{Y}) = \mathcal{N}(\mu_{\text{post}}(\mathbf{X}, \mathbf{Y}), \text{diag}(\sigma_{\text{post}}(\mathbf{X}, \mathbf{Y})))\label{e3}
\end{equation}

In the training process, we compute the loss as shown in Eq.\ref{e4}. Given the raw image $X$ and the ground truth segmentation $Y$, $S$ is the predicted segmentation. The Dice-CE loss, treating the output $S$ as the parameterization of a pixel-wise categorical distribution $D$, penalizes differences between $S$ and $Y$. We empirically set the scaling coefficient $\beta$ to 1.
\begin{equation}
\begin{split}
\mathcal{L} = {E}_{z\sim Q\left ( \cdot \mid Y, X \right )} \left[ - \log_{}{{D}} \left( Y \mid S\left( X, z \right) \right) \right] \\
+ \beta \times {D}_{KL} \left( Q\left( z \mid Y, X \right) \parallel P\left( z \mid X \right) \right)
\end{split}
\label{e4}
\end{equation}

\subsection{Uncertainty-aware Adapter}
The multi-expert knowledge latent distribution $\mathcal{M}e(\cdot)$ simulates the collective decision-making process of multiple experts. Given any input image $X$, the distribution can generate unlimited possible low-dimensional mappings (stochastic samples $z$) that correspond to different plausible segmentations. The stochastic sample $z$ characterizes the representation of uncertainty inherent in ambiguous medical images. The key innovation of Uncertainty-aware Adapter is designed to help SAM effectively integrate these uncertainty samples, ensuring the model can generate appropriate responses for different uncertainty representations. However, it is worth noting that adapters at different positions actually process image embeddings with different features, such as image structures, textures, and semantic information. Uniformly sampling a single annotation latent representation $z$ for different adapters leads to a feature mismatch, which is demonstrated by our subsequent experiments. It is necessary to align $z$ with the visual representations that the adapters are processing. Therefore, we propose a lightweight module called Position-conditioned Attention (Poscon-Att) to align the uncertainty samples $z$ with image embeddings.

\noindent
\textbf{Position-conditioned Attention (Poscon-Att)}:  
Direct computation between the high-dimensional visual features $E$ and the stochastic variable $z$ can be computationally expensive. Instead, we utilize interactions with low-dimensional positional encodings, allowing the adapter to perceive uncertainty while avoiding additional parameter overhead during fine-tuning. As illustrated in Fig.\ref{fig3}, the module takes as input the positional variable ${p}_{i}$ and uncertainty sample $z$, then outputs the position-conditioned uncertainty feature $G(z)$ and positional variable ${p}_{i+1}$ for the next adapter. Specifically, we first use global average pooling and a two-layer MLP to map the image embeddings $E_{0}$, encoded by the $0$-th Transformer block, to the positional encoding ${p}_{0}$ (${p}_{i}\in {\mathbb{R}}^{1\times L}$, where $L$ is the number of adapters). Subsequently, we employ scaled dot-product attention following \cite{vaswani2017attention} to integrate positional features and stochastic sample features, generating the latent variable $Z'$. We use different linear projections to map the positional encoding ${p}_{i}$ and the stochastic sample $z$ to features $P_{K}$, $P_{V}$, $Z_{Q}$, and $Z_{V}$ respectively (where they have the same shape as $z\in {\mathbb{R}}^{B\times C}$, where $B$ is the batch size and $C$ is the latent space dimension). The attention mechanism is used to align uncertainty sampling features to the adapters' positions as illustrated in Eq.\ref{e5}.
\begin{equation}  
\text{\( Z' \)} = \text{softmax}\left( \frac{Z_Q P_K^\top  }{ \sqrt{C} } \right) Z_V\label{e5}
\end{equation}

Conditioned uncertainty representation $Z'$, which contains the image semantic information and matches the stages of feature processing, undergoes channel interaction to obtain the position-conditioned uncertainty feature $G(z)$. More specifically, we use element-wise multiplication ($\odot$) between $Z'$ and $P_{V}$ to further integrate position information into the uncertainty feature. Then, we reshape the uncertainty feature to match the dimension of the image embedding. We transform feature tensors from ${\mathbb{R}}^{B\times C}$ to ${\mathbb{R}}^{B\times H\times W\times \text{dim}}$ using channel replication and a three-layer MLP, where $H$ and $W$ denote the height and width of patch embeddings processed by the image encoder, and $\text{dim}$ is the channel dimension of patch embeddings. Finally, we apply reshape and MLP operations to the positional encoding features $P_{V}$ to adapt them for the subsequent adapter feature processing stage, thereby representing the updated positional encoding features $p_{i+1}$ for the next adapter. The calculation of $G(z)$ is as follows:
\begin{equation}  
\text{\( G(z) \)} = \text{MLP}(\text{Repeat}(Z^{'}\odot P_{V}, H or W, dim=1,2), dim=3)\label{e6}
\end{equation}

\noindent
\textbf{Prompt Channel Attention (Prompt-CAtt)}: Considering the significant impact of interactive prompts on different ambiguous segmentation tasks, we employ prompt-conditioned adaptation to enhance the interaction between visual prompts and the adapter, as illustrated in Fig.\ref{fig3}. Toward that end, we introduce a Prompt Channel Attention (Prompt-CAtt) module. We find that directly interacting with high-level image embeddings causes low-level semantic prompt embeddings to be overshadowed, as demonstrated by our ablation studies. To address this, we apply channel attention in Prompt-CAtt to establish a strong connection between uncertainty and visual prompts, which enhances UA-SAM's robustness across various ambiguous medical image segmentation tasks involving different visual prompts. Specifically, we use a global max pooling to capture high activations from the user prompt embeddings and an average pooling to obtain a smooth estimation of the intended position. We then integrate an MLP to map these weights to the same dimension as $G(z)$, and perform a conditional enhancement on $G(z)$ using channel-wise dot product. Finally, we sum the features obtained from the two branches to generate uncertainty features that are effectively constrained by the visual prompt information. In this way, user prompt information is used to condition the uncertainty features, enabling the model to produce diverse outputs while maintaining robust prompt adaptability.

\begin{table*}[t]
\centering
\caption{Quantitative comparison with state-of-the-art methods on LIDC-IDRI and REFUGE-cup datasets. Best results are in bold. For all the compared methods, the results are obtained either through reproduction using their official implementations or directly from the experimental results reported in their respective papers.}
\label{tab1}
\setlength{\tabcolsep}{6pt} 
\renewcommand{\arraystretch}{1.1} 
\small 
\begin{tabular}{lccc|ccc|ccc}
\hline
\multirow{2}{*}{Category} & \multirow{2}{*}{Method} & \multirow{2}{*}{\shortstack{SAM-\\based}} & \multirow{2}{*}{\shortstack{Trainable\\ Params (M)}} & \multicolumn{3}{c|}{LIDC-IDRI} & \multicolumn{3}{c}{REFUGE-cup} \\
\cline{5-10}
& & & & Dice$\uparrow$ & mIoU$\uparrow$ & GED$\downarrow$ & Dice$\uparrow$ & mIoU$\uparrow$ & GED$\downarrow$ \\
\hline
\multirow{5}{*}{\shortstack{Deterministic\\Methods}} 
& U-Net & & 7.76 & 0.690 & 0.624 & 0.415 & 0.789 & 0.676 & 0.475 \\
& SAM-adapter & \checkmark & 4.48 & 0.876 & 0.781 & 0.345 & 0.884 & 0.768 & 0.371 \\
& Med-SA & \checkmark & 7.24 & 0.869 & 0.773 & 0.333 & 0.889 & 0.771 & 0.433 \\
& MedSAM & \checkmark & 97.29 & 0.849 & 0.769 & 0.367 & 0.864 & 0.758 & 0.385 \\
& MedSAM2 & \checkmark & 37.14 & 0.852 & 0.779 & 0.311 & 0.847 & 0.741 & 0.367 \\
\hline
\multirow{6}{*}{\shortstack{Uncertainty\\Methods}} 
& SAM-U & \checkmark & 0 & 0.615 & 0.421 & 0.656 & 0.632 & 0.446 & 0.610 \\
& MRNet & & 81.19 & 0.707 & 0.624 & 0.504 & 0.849 & 0.703 & 0.440 \\
& LS-UNet & & 29.94 & 0.623 & 0.578 & 0.540 & 0.755 & 0.631 & 0.548 \\
& Ensemble U-Net & & 23.28 & 0.638 & 0.589 & 0.475 & 0.744 & 0.652 & 0.459 \\
& Prob UNet & & 10.34 & 0.602 & 0.513 & 0.331 & 0.682 & 0.555 & 0.559 \\
& CCDM & & 30 & - & - & 0.239 & - & - & - \\
& MedSegDiff & & 25 & 0.717 & 0.602 & 0.420 & 0.859 & 0.762 & 0.279 \\
\rowcolor[HTML]{E8E8E8} 
& \textbf{UA-SAM (Ours)} & \checkmark & 8.04 & \textbf{0.889} & \textbf{0.786} & \textbf{0.209} & \textbf{0.897} & \textbf{0.789} & \textbf{0.230} \\
\hline
\end{tabular}
\end{table*}

\section{Experiments}
\subsection{Datasets}
To verify the effectiveness of the proposed framework, we  evaluated it on three publicly available multi-expert annotations datasets: lung nodule segmentation dataset LIDC-IDRI \cite{clark2013cancer}, optic-cup segmentation dataset REFUGE2-cup \cite{fang2022refuge2}, and multiple organs segmentation dataset QUBIQ \cite{menzequantification}. These datasets encompass seven multi-annotated benchmarks and include data from various image modalities, such as color fundus images, CT scans, and MRI.

\noindent
\textbf{LIDC-IDRI}. The LIDC-IDRI dataset is a lung nodule segmentation dataset with lesion areas annotated by four radiologists. The dataset includes 15,096 thoracic CT images and cropped $128 \times 128$ patches around the lesions. We divide the dataset into a training set and a testing set at an 80:20 ratio. The training set contains 12,076 images, while the testing set consists of 3,018 images.

\noindent
\textbf{REFUGE2}. We conduct experiments on the REFUGE2 optic-cup dataset, consisting of 1,200 fundus images, each annotated by seven radiologists. Training, validation, and test sets each contain 400 images. This dataset is provided for glaucoma analysis. Each image is high-resolution, sized at 2056 × 2124. To focus more on the segmentation of blurred regions, we crop each image to 384 × 384 around the optic-cup center. 

\noindent
\textbf{QUBIQ}. QUBIQ is used for quantifying uncertainties in biomedical images. It contains four different segmentation datasets with CT and MRI modalities, including brain-growth (one task, MRI, seven raters, 34 cases for training and 5 cases for testing), brain-tumor (one task, MRI, three raters, 28 cases for training and 4 cases for testing), prostate (two subtasks, MRI, six raters, 48 cases for training and 7 cases for testing), kidney (one task, CT, three raters, 20 cases for training and 4 cases for testing).

\subsection{Experimental Settings}
\subsubsection{Training details} During training, we employed the Adam optimizer with an initial learning rate of $1e^{-4}$ and a StepLR scheduler for learning rate decay. For each image, labels were randomly sampled from the available multiple annotations. All images from REFUGE2 are cropped around the center of the cup to a size of 384 × 384 pixels, while others use their default sizes. To ensure fair comparison, we used \textbf{SAM-ViT/B} as the backbone for all SAM variants and set the \texttt{multimask\_output} parameter to 1. All baseline methods were configured with their default settings as reported in the original papers. The experiments were implemented in PyTorch 2.3.0 and trained on two NVIDIA RTX 3090 GPUs with 24 GB memory. All adapter-based models were trained for up to 50 epochs with early stopping based on validation loss.

\subsubsection{Evaluation details} 
UA-SAM simulates the multi-expert annotation paradigm. It generates diverse predictions that not only align with real physician annotation distributions but also match multi-expert consensus. To comprehensively assess our method, we use two categories of metrics. For consensus evaluation, we apply a majority voting strategy across multiple annotated labels to derive the consensus mask. We measure Dice score and mIoU to quantify segmentation accuracy and overlap. For distribution evaluation, we compute the Generalized Energy Distance (GED) to assess the alignment between predicted and ground truth annotation distributions. GED is calculated using 16 sampled predictions per image. All metrics are reported as mean values across multiple runs. The calculation methods for these metrics are as follows:

\noindent
\textbf{Dice Score:} The Dice similarity coefficient measures the overlap between predicted and ground truth segmentations, computed using the MONAI library with the formula:
\begin{equation}
\text{Dice} = \frac{2 \times |P \cap G|}{|P| + |G|}
\end{equation}
where $P$ represents the predicted mask and $G$ denotes the ground truth mask. 

\noindent
\textbf{Mean Intersection over Union (mIoU):} The IoU metric evaluates pixel-wise accuracy by calculating the ratio of intersection to union between predicted and ground truth masks:
\begin{equation}
\text{IoU} = \frac{|P \cap G|}{|P \cup G|}
\end{equation}
We compute the mean IoU across all images in each batch, with a small epsilon value ($1e^{-8}$) added to prevent division by zero.

\noindent
\textbf{Generalized Energy Distance (GED):} As a distributional metric, GED quantifies the similarity between the model's predicted distribution and the true multi-expert annotation distribution. It measures the discrepancy between two probability distributions using:
\begin{equation}
\text{GED} = 2\mathbb{E}[d(X,Y)] - \mathbb{E}[d(X,X')] - \mathbb{E}[d(Y,Y')]
\end{equation}
where $X$ and $Y$ represent samples from the predicted and ground truth distributions respectively, and $d(\cdot,\cdot)$ denotes the distance function based on Intersection over Union. Lower GED values indicate better alignment between predicted and true annotation distributions.

\subsubsection{Prompt Synthesizing Strategy} For training, we randomly selected one point from the ground truth region as the prompt. For evaluation, a majority voting strategy was applied across multiple annotated labels to generate the final mask. SAM variants were evaluated using varying numbers of points and bounding boxes as prompts, both derived from the majority voting masks. Points were randomly sampled from the ground truth region, while bounding boxes \cite{cheng2023sam} were defined as the maximum enclosing rectangles with each coordinate randomly perturbed by up to 5 pixels.

\begin{figure*}[!t]
  \centering
 \includegraphics[width=0.9\linewidth]{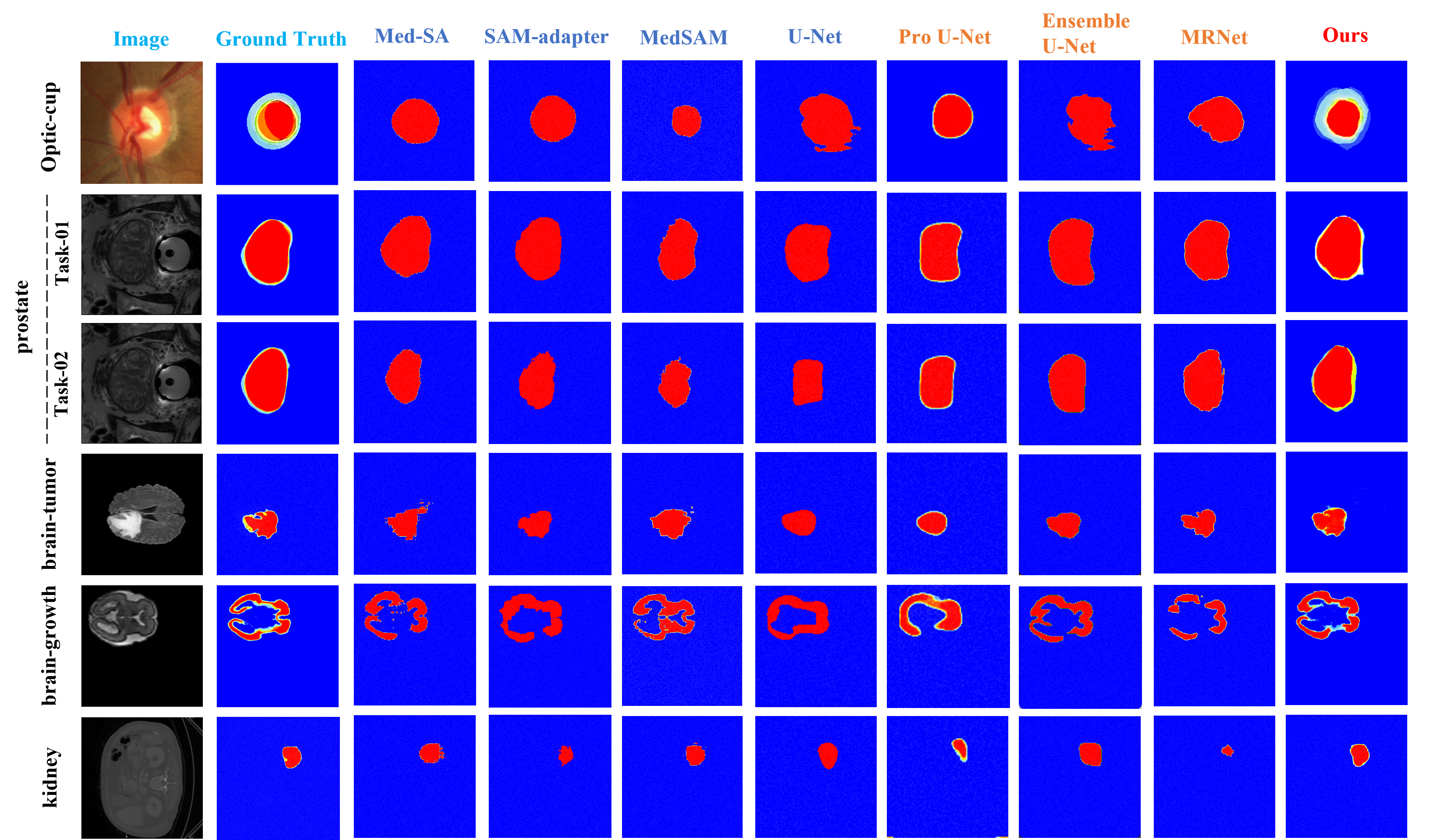}
  \caption{UA-SAM shows superior segmentation visualization. Our results exhibit more prominent uncertainty boundaries while maintaining shapes that are closer to the ground truth. Here, the Ensemble U-Net is implemented by training UNet under three different random seeds. Due to the large discrepancy in prediction shapes, we employed majority voting to aggregate the results from multiple U-Nets. }\label{fig4}
\end{figure*}

\subsection{Main Results}
\subsubsection{Quantitative assessment} We compared our UA-SAM model with state-of-the-art (SOTA) segmentation methods, classified into deterministic methods and uncertainty methods, with quantitative results in Table~\ref{tab1}. The deterministic methods encompass typical adapter-based SAM fine-tuning approaches, including SAM-adapter (which inserts the original adapters at the same locations as UA-SAM) and Med-SA \cite{wu2023medical}, generalized SAM fine-tuning methods such as MedSAM \cite{ma2024segment} and MedSAM2 \cite{zhu2024medical}, as well as the traditional U-Net \cite{ronneberger2015u}. Among the uncertainty methods, we compared our approach with the interactive uncertainty method SAM-U \cite{deng2023sam}, widely adopted uncertainty-aware approaches including Prob UNet \cite{kohl2018probabilistic}, Ensemble U-Net \cite{lakshminarayanan2017simple}, LS-UNet \cite{jensen2019improving}, and MRNet \cite{ji2021learning}, as well as diffusion-based generative approaches CCDM \cite{zbinden2023stochastic} and MedSegDiff \cite{wu2024medsegdiff}.

\begin{table}[ht]
\footnotesize
\centering
\setlength{\tabcolsep}{5pt} 
\caption{Quantitative evaluation of the QUBIQ benchmark by Dice score, including segmentation of braintumor $D_{tum}$, braingrowth $D_{brn}$, two prostate tasks $D_{pr1}$ and $D_{pr2}$, and kidney $D_{kid}$. 1-point prompt is employed for SAM variants. Best results are in bold. Certain methods could not be successfully reproduced on the QUBIQ dataset owing to convergence difficulties, and consequently their results are omitted from the comparison.}
\label{tab2}
\renewcommand{\arraystretch}{1.1}
\begin{tabular}{l|ccccc|c}
\hline
Method & $D_{tum}$ & $D_{brn}$ & $D_{pr1}$ & $D_{pr2}$ & $D_{kid}$ & Avg \\
\hline
U-Net & 0.895 & 0.729 & 0.841 & 0.695 & 0.824 & 0.796 \\
MedSAM & 0.826 & 0.761 & 0.898 & 0.683 & 0.862 & 0.806 \\
SAM-adapter & 0.824 & 0.772 & 0.935 & 0.795 & 0.851 & 0.836 \\
Med-SA & 0.843 & 0.802 & 0.911 & 0.771 & 0.723 & 0.810 \\
SAM-U & 0.908 & 0.817 & 0.846 & 0.725 & 0.895 & 0.838 \\
Ensemble U-Net & 0.901 & 0.741 & 0.868 & 0.746 & 0.893 & 0.829 \\
Prob UNet & 0.891 & 0.645 & 0.688 & 0.653 & 0.606 & 0.696 \\
LS-UNet & 0.858 & 0.827 & 0.862 & 0.690 & 0.723 & 0.792 \\
MRNet & 0.890 & 0.822 & 0.873 & 0.760 & 0.581 & 0.785 \\
\rowcolor[HTML]{E8E8E8} 
\textbf{UA-SAM} & \textbf{0.911} & \textbf{0.840} & \textbf{0.938} & \textbf{0.818} & \textbf{0.947} & \textbf{0.891} \\
\hline
\end{tabular}
\end{table}

For SAM variants, we report the optimal results across different prompt configurations (1-point, 2-point, 3-point, and bounding box prompts), with GED calculated using a 1-point prompt. Under bounding box prompt conditions, our model demonstrated superior performance, achieving Dice scores of 0.889 and 0.897, along with mIoU scores of 0.786 and 0.789 on the LIDC-IDRI and REFUGE2 datasets, respectively. UA-SAM also achieves the lowest GED values of 0.209 and 0.230, indicating that its predictions are closer to the distribution of real human doctors. We adapt SAM using a multi-expert collective intelligence paradigm, enabling SAM to capture rich information from different expert annotations, thereby producing outputs that better align with expert group consensus, and the output distribution more closely approximates multi-expert annotations. Additionally, as reported in Table~\ref{tab2}, quantitative comparisons on QUBIQ further confirm that the Uncertainty-aware Adapter exhibits superior generalization capabilities compared to a standard adapter when fine-tuning SAM for segmenting ambiguous medical images, even outperforming other specialized uncertainty methods. 

Furthermore, we present a comparison between SAM variants and UA-SAM under different prompt settings in Table~\ref{tab3}. Our UA-SAM outperforms others with the highest Dice score in almost all prompt settings. Even in optic cup segmentation, UA-SAM surpasses other variants that employ multi-point prompts by using only single-point prompts. This confirms the necessity of incorporating visual prompt information to constrain the adaptation process in UA-SAM, as it enhances prompt sensitivity and yields more robust segmentation results for ambiguous medical images.

\begin{table}[ht]
\centering
\caption{Performance comparison of different SAM variants with various prompt types. Best results are in bold.}
\label{tab3}
\setlength{\tabcolsep}{8pt}
\renewcommand{\arraystretch}{1.1}
\begin{tabular}{lllll}
\hline
\multicolumn{5}{c}{LIDC-IDRI}              \\ \hline
Prompts         & 1-point             & 2-points             & 3-points             & boxes          \\
SAM-Adapter     & 0.861          & 0.651          & 0.628          & 0.876          \\
Med-SA          & 0.869          & 0.799          & 0.825          & 0.868          \\
MedSAM          & 0.849          & 0.832          & \textbf{0.829} & 0.829          \\
\rowcolor[HTML]{E8E8E8} 
\textbf{UA-SAM} & \textbf{0.876} & \textbf{0.832} & 0.746          & \textbf{0.889} \\ \hline
\multicolumn{5}{c}{REFUGE2-cup}            \\ \hline
Prompts         & 1-point             & 2-points             & 3-points             & boxes          \\
SAM-Adapter     & 0.823          & 0.842          & 0.845          & 0.884          \\
Med-SA          & 0.830          & 0.844          & 0.848          & 0.889          \\
MedSAM          & 0.821          & 0.852          & 0.864          & 0.808          \\
\rowcolor[HTML]{E8E8E8} 
\textbf{UA-SAM} & \textbf{0.864} & \textbf{0.867} & \textbf{0.870} & \textbf{0.897} \\ \hline
\end{tabular}
\end{table}

\subsubsection{Qualitative assessment}
Apart from superior performance, our most significant advantage is the capability to produce credible segmentation margins for ambiguous tissues or organs, as demonstrated in Fig.~\ref{fig4}, thereby extending its clinical utility. Fig.~\ref{fig4} reveals that conventional adapter-based methods, such as Med-SA and SAM-Adapter, yield deterministic predictions for ambiguous targets, limiting their adoption in clinical practice. Other uncertainty methods like Prob U-Net and MRNet produce uncertainty maps that are both smaller and morphologically less aligned with the ground-truth labels. In contrast, our approach delivers well-defined uncertainty regions whose contours closely match the reference annotations. Qualitative results demonstrate that our approach successfully enables the paradigm shift in SAM adaptation from single-expert deterministic outputs to multi-expert collective intelligence representation.

\begin{table}[t]
\small
\centering
\setlength{\tabcolsep}{5pt}
\caption{Ablation study results by Dice score $D_{lung}$ of LIDC-IDRI and $D_{cup}$ of optic-cup segmentation. Best results are in bold. Note that '\textit{z} and \textit{p}' represents sample \textit{z} without being conditioned by \textit{p} and means \textit{z} directly concatenated with \textit{p}.}
\label{tab4}
\renewcommand{\arraystretch}{1.1}
\begin{tabular}{cccccc|cc}
\hline
\multirow{2}{*}{\textit{z}} & \multirow{2}{*}{\textit{p}} & \textit{z} and & Poscon- & Prompt- & Prompt- & \multirow{2}{*}{$D_{lung}$} & \multirow{2}{*}{$D_{cup}$} \\
 &  & \textit{p} & Att & add & CAtt &  &  \\
\hline
\checkmark &  &  &  &  &  & 0.860 & 0.823 \\
 & \checkmark &  &  &  &  & 0.862 & 0.826 \\
\checkmark & \checkmark & \checkmark &  &  &  & 0.865 & 0.828 \\
\checkmark & \checkmark &  & \checkmark &  &  & 0.872 & 0.847 \\
\checkmark & \checkmark &  & \checkmark & \checkmark &   & 0.868 & 0.848 \\ 
\checkmark & \checkmark &  & \checkmark & & \checkmark & \textbf{0.876} & \textbf{0.864} \\ 
\hline
\end{tabular}
\end{table}

\begin{table*}
\centering
\caption{Ablation study on QUBIQ by Dice score, including segmentation of braintumor $D_{tum}$, braingrowth $D_{brn}$, two prostate tasks $D_{pr1}$ and $D_{pr2}$, and kidney $D_{kid}$. 1-point prompt is employed for SAM variants. Best results are in bold. Note the '\textit{z} and \textit{p}' represents sample \textit{z} without being conditioned by \textit{p} and means \textit{z} directly concatenated with \textit{p}.}\label{tab5}
\renewcommand{\arraystretch}{1.1}
\begin{tabular}{ccccc|ccccc} 
\hline
\begin{tabular}[c]{@{}c@{}}Uncertainty \\ sample \textit{z}\end{tabular} & 
\begin{tabular}[c]{@{}c@{}}Position \\ variant \textit{p}\end{tabular} & 
\textit{z} and \textit{p} & 
Poscon-Att & 
Prompt-CAtt & 
$D_{tum}$ & 
$D_{brn}$ & 
$D_{pr1}$ & 
$D_{pr2}$ & 
$D_{kid}$ \\ \hline
\checkmark &  &  &  &  & 0.889 & 0.812 & 0.878 & 0.807 & 0.848 \\
 & \checkmark &  &  &  & 0.869 & 0.791 & 0.911 & 0.797 & 0.811 \\
\checkmark & \checkmark & \checkmark &  &  & 0.815 & 0.783 & 0.880 & 0.727 & 0.715 \\
\checkmark & \checkmark &  & \checkmark &  & 0.904 & 0.824 & 0.928 & 0.813 & 0.938 \\
\checkmark & \checkmark &  & \checkmark & \checkmark & \textbf{0.911} & \textbf{0.840} & \textbf{0.938} & \textbf{0.818} & \textbf{0.947} \\ \hline
\end{tabular}%
\end{table*}

\subsubsection{Ablation study}
We conducted comprehensive ablation studies to validate the effectiveness of each component proposed in UA-SAM. In Tables~\ref{tab4} and \ref{tab5}, we compared several methods based on SAM, including separately concatenating the stochastic sample $z$ and the learnable positional variable $p$ into the adapter, as well as combining both through different interaction methods. We employed a 1-point prompt for all models. The results reveal no significant changes in model performance for most tasks when directly concatenating the sample $z$ into the adapter, which is similar to Prob U-Net. As previously mentioned, we utilize Poscon-Att to condition uncertainty sampling and achieve feature matching with the adapter, resulting in a significant performance improvement. Furthermore, we observed that during prompt adaptation, simple fusion methods like direct addition of prompt embeddings to high-level visual features lead to either performance degradation or no improvement. This further validates the effectiveness of the proposed Prompt-CAtt.

Additionally, we conducted comprehensive experiments on the dimensionality of the latent distribution and the number of adapters to identify the optimal hyper-parameters, ensuring the best performance of UA-SAM. As illustrated in Table~\ref{tab6}, in Sub-table (a), we observe that the optimal performance is achieved when the dimension $C$ of the multi-expert knowledge latent distribution $\mathcal{M}e(\cdot)$ is set to 6, indicating that at this point the latent space best captures the annotation patterns of the experts. Sub-table (b) confirms that increasing the number of inserted adapter layers steadily improves model performance, and equipping more adapters yields consistently stronger results.

\begin{figure}[t]
    \centering    \includegraphics[width=0.95\linewidth]{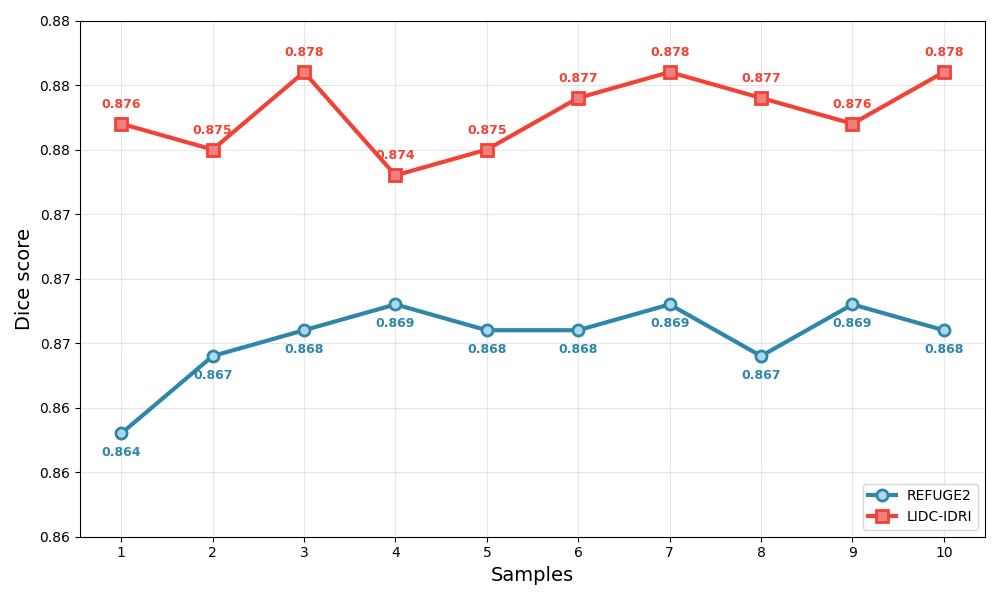}
    \caption{The impact of sampling times. For each image, the model predicts $n$ times and then aggregates the results into the final mask via majority voting.}
    \label{fig5}
\end{figure}

Previously, all evaluations used single sampling for each input image. Here, we analyze the performance change when using multiple samples for a single input and then aggregating the samples using a majority voting strategy, as shown in Fig.~\ref{fig5}. As the number of samples increases, UA-SAM's performance shows only slight improvement. This is because, although the multiple sampled predictions are diverse, their distribution is close to the results collectively agreed upon by multiple expert annotators in the labels; they all contain common regions, so the result after majority voting does not change much. The model's performance remains stable in representing the multi-expert consensus, so no further sampling is needed to improve performance, only to enhance diversity.

\begin{table}[]
\centering
\caption{Ablation on different hyperparameter settings}\label{tab6}
\setlength{\tabcolsep}{9pt} 
\renewcommand{\arraystretch}{1.1}
\begin{tabular}{ccccccc}
\hline
\multicolumn{4}{c}{(a) Latent space dimension} & & \multicolumn{2}{c}{(b) Adapter inserted layers} \\ \cline{1-4} \cline{6-7}
$C$ & \(D_{cup}\) & $C$ & \(D_{cup}\) & & Layers & \(D_{cup}\) \\ \cline{1-4} \cline{6-7}
2 & 0.822 & \textbf{6} & \textbf{0.864} & & 1--3 & 0.803 \\
3 & 0.818 & 7 & 0.820 & & 1--6 & 0.817 \\
4 & 0.817 & 8 & 0.811 & & 1--9 & 0.854 \\
5 & 0.819 & 9 & 0.820 & & \textbf{1--12} & \textbf{0.864} \\ \hline
\end{tabular}
\end{table}

\begin{figure}
  \centering
\includegraphics[width=\linewidth]{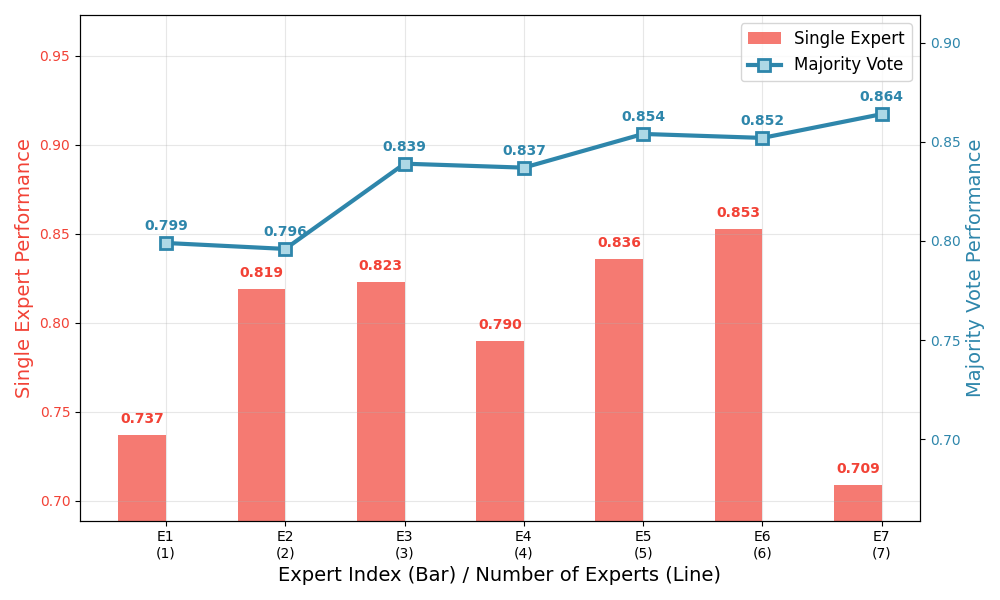}
  \caption{Consensus matching. Evaluation performance rises as the number of experts involved in evaluation increases.}\label{fig6}
\end{figure}

\begin{figure}
  \centering
\includegraphics[width=0.95\linewidth]{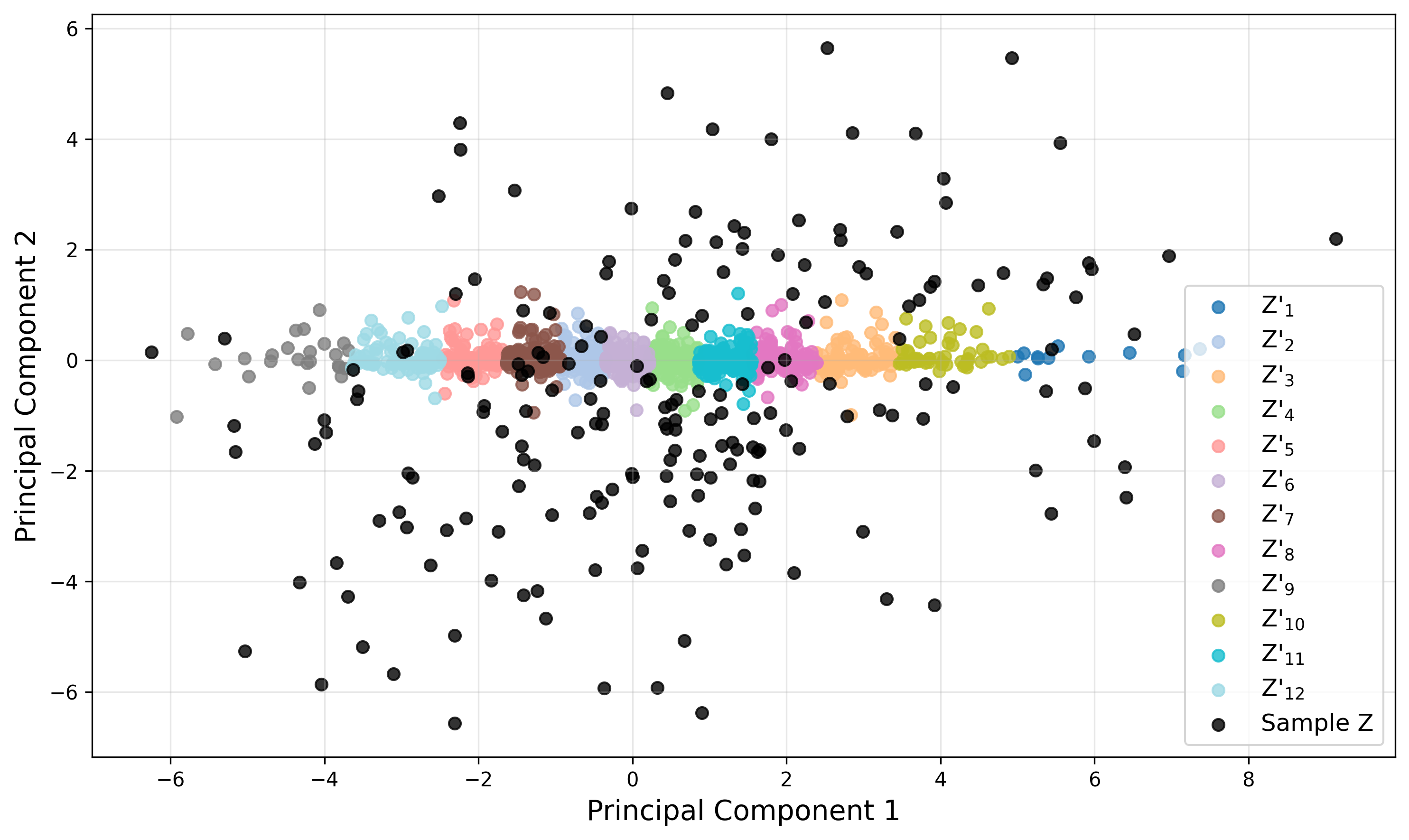}
  \caption{The visualization of $Z'$ and sample $z$ on the REFUGE2-cup. PCA and K-means for clustering.}\label{fig7}
\end{figure}

\begin{table}[ht]
\centering
\setlength{\tabcolsep}{5pt}
\caption{Performance comparison of SAM-based methods including trainable parameters (Params, M), inference time (Infer, ms/img), frames per second (FPS, imgs/s), Dice score of LIDC-IDRI and optic-cup segmentation under 1-point prompt.}
\label{tab7}
\renewcommand{\arraystretch}{1.1}
\begin{tabular}{lccccc}
\hline
Method & \shortstack{Params} & \shortstack{Infer} & \shortstack{FPS} & \shortstack{$D_{lung}$} & \shortstack{$D_{cup}$} \\
\hline
MedSAM & 97.29 & 84.99 & 11.77 & 0.849 & 0.821 \\
Med-SA & 7.24 & 87.57 & 11.42 & 0.869 & 0.830 \\
SAM-adapter & 4.48 & 91.55 & 10.92 & 0.861 & 0.823 \\
\textbf{UA-SAM} & 8.04 & 112.95 & 8.85 & \textbf{0.876} & \textbf{0.864} \\
\hline
\end{tabular}
\end{table}

\subsubsection{Additional analysis} 
Since we incrementally modified the adapters for enabling SAM to align with the collective intelligence paradigm, we conducted a comprehensive analysis to evaluate both the parameter efficiency of our Uncertainty-aware Adapter and its computational impact on model inference speed.
As shown in Table \ref{tab7}, our method increases trainable parameters moderately compared to other adapter-based approaches, but remains practical for deployment. Although inference speed slightly decreases due to the additional computation in adapters, the performance improvements across all metrics justify this minor cost. The computational trade-off is minimal when considering the significant enhancement in clinical applicability and diagnostic reliability that our method provides.
In addition, we evaluated whether UA-SAM can effectively capture multi-expert consensus, as illustrated in Fig.\ref{fig6}. Consensus masks were generated with varying numbers of expert annotations. As the number of experts involved in the evaluation increases, the model’s performance shows an upward trend, indicating that UA-SAM can effectively accommodate clinical scenarios with multiple experts and accurately extract consensus information from divergent opinions.
Furthermore, the Fig.\ref{fig7} shows that $Z'$ clusters differently after integrating the position encoding into the adapters, thereby verifying the effectiveness of our Poscon-Att design in conditioning uncertainty samples to align with the appropriate stage of feature processing within the adapters. This further supports our motivation.

\begin{figure}[htp]
  \centering
\includegraphics[width=0.95\linewidth]{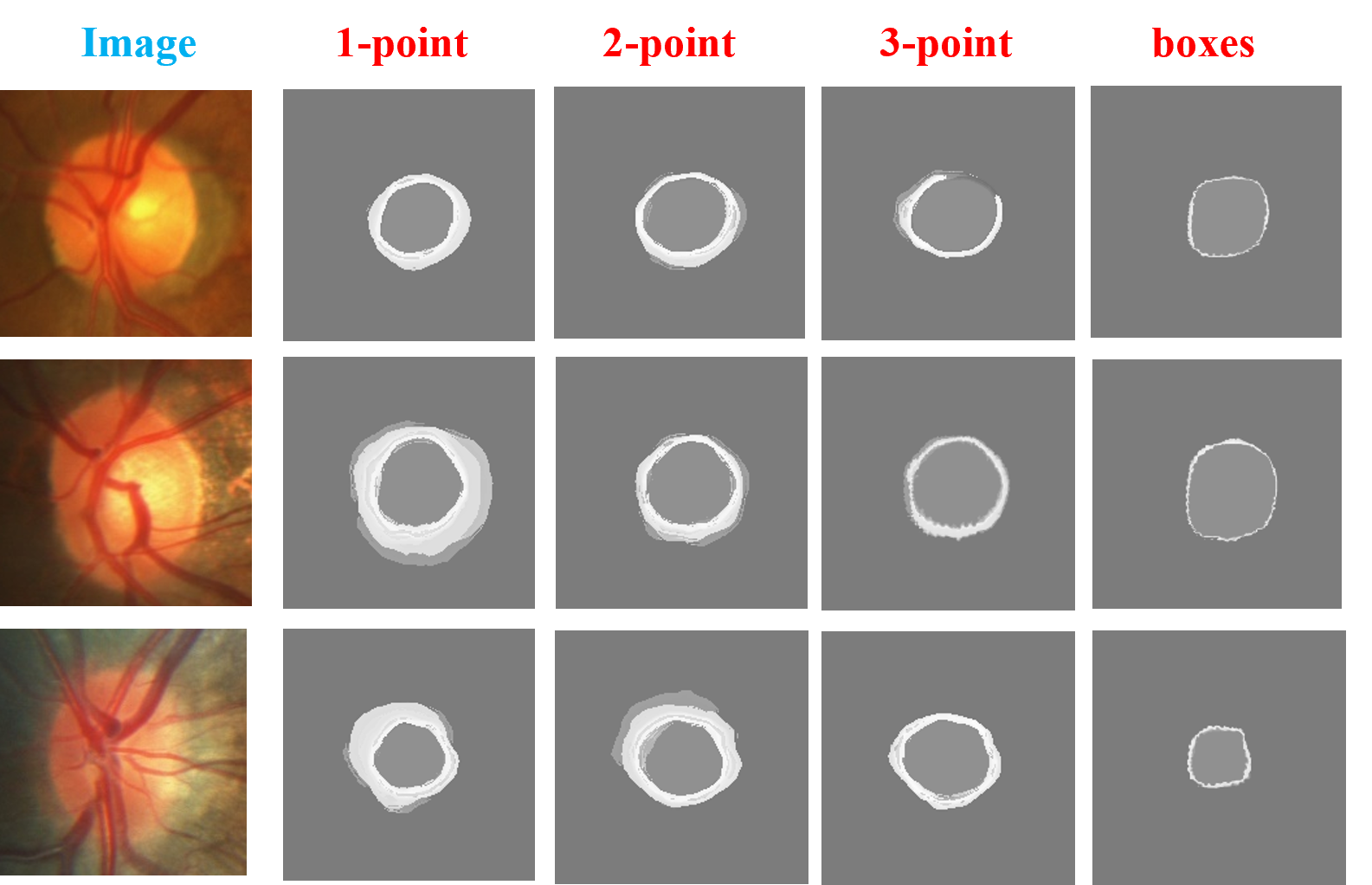}
  \caption{Different prompts on the REFUGE2-cup. Each result is based on eight stochastic samples. Gray areas mark regions where all samples agree perfectly and white areas highlight zones of high prediction uncertainty. Our method supports controllable multi-annotation patterns. By changing the visual prompts, we can control the level of uncertainty in the outputs.}\label{fig8}
\end{figure}
To verify the robustness of UA-SAM across prompt variations, we systematically varied the prompts and visualized their impact on output uncertainty in Fig.~\ref{fig8}.
As the number of points increases, the outputs' uncertainty gradually decreases (indicated by the white areas), tending towards more deterministic predictions. When using a bounding box as the prompt, the diversity of the model outputs is extremely low. These results indicate that UA-SAM exhibits robust performance across different prompts, further validating its prompt-adaptive capability. It also highlights a significant advantage of our approach compared to previous non-interactive uncertainty methods: our method allows users to easily adjust the degree of uncertainty in the results through different prompts, making it adaptable to a variety of clinical scenarios.

\section{Conclusion}
We propose an Uncertainty-aware Adapter, a multi-expert collective intelligence adaptation paradigm for adapting SAM to ambiguous medical segmentation. This paradigm shift from single-expert predictions to collective intelligence representation aligns with clinical practice where multiple specialists provide diverse yet valid interpretations. Evaluations on seven medical benchmarks demonstrate superior performance while enabling multiple plausible segmentation hypotheses to be generated. Our method makes SAM more reliable and clinically applicable for downstream medical tasks.

\bibliographystyle{ieeetr}
\bibliography{tmi.bib}

\end{document}